\documentclass[a4paper,11pt]{article}
\usepackage{pos}
\usepackage{layouts}
\usepackage{wrapfig}
\usepackage{floatrow}

\title{Designing weight regularizations based on Lefschetz thimbles to stabilize complex Langevin}
\ShortTitle{Designing weight regularizations}

\author[a]{Kirill Boguslavski}
\author*[a]{Paul Hotzy}
\author[a]{David I.\ M\"uller}

\affiliation[a]{Institute of Theoretical Physics, TU Wien, Wiedner Hauptstraße 8-10, 1040 Vienna, Austria}

\emailAdd{kirill.boguslavski@tuwien.ac.at}
\emailAdd{paul.hotzy@tuwien.ac.at}
\emailAdd{dmueller@hep.itp.tuwien.ac.at}

\abstract{The complex Langevin (CL) method shows significant potential in addressing the numerical sign
	problem. Nonetheless, it often produces incorrect results when used without any stabilization techniques. Leveraging insights from previous research that links Lefschetz thimbles and CL, we explore
	a strategy to regularize the CL method to address this issue of incorrect convergence. Specifically,
	we implement weight regularizations inspired by the associated Lefschetz thimble structure and correct the bias to retrieve the correct results of the original theory. We demonstrate the effectiveness of this approach by solving the SU(N) Polyakov chain model and various scalar models, including the cosine model and the one-link model, across a broad range of couplings where the CL method previously failed. We also discuss the potential application of these insights to gauge theories in practical scenarios.
}

\FullConference{The 41st International Symposium on Lattice Field Theory (LATTICE2024)\\
	28 July - 3 August 2024\\
	Liverpool, UK\\}

\newcommand{\pToFigs}{figures}

\begin{document}
	\maketitle
	
	\section{Introduction}
	
	Lattice field theory provides a robust framework for non-perturbative studies in high-energy and condensed-matter physics, with Monte Carlo (MC) simulations serving as its cornerstone. While highly effective for systems in thermal equilibrium, where the path integral weight is positive-definite, traditional MC methods fail when faced with complex or non-positive weight functions. Such situations arise in systems at finite density or involving real-time dynamics, leading to the well-known numerical sign problem. This longstanding computational challenge renders many physically relevant systems inaccessible to direct lattice calculations.
	
	Several alternative approaches have been proposed to circumvent the sign problem, including the complex Langevin (CL) method \cite{Parisi:1983mgm}. In CL, the degrees of freedom are analytically continued to the complex plane and the path integral is reformulated in terms of a complex stochastic process. While computationally efficient and scalable, the method's convergence properties often depend on the underlying structure of the system and can fail under certain conditions. These failures are frequently linked to the interplay between CL dynamics and the Lefschetz thimbles \cite{Cristoforetti:2012su} — complex integration contours defined by the critical points of the action \cite{Aarts:2014nxa}.
	
	In these proceedings, we investigate the connection between Lefschetz thimbles and the convergence properties of CL, focusing on systems characterized by one dominant compact thimble. For such systems, CL reliably reproduces correct results, which aligns with conjectures made in the literature \cite{Salcedo:2016kyy}. To probe this behavior, we employ additive weight regularization to achieve the desired thimble structure and correct the resulting bias in a series of toy models, including the complex cosine model and the Polyakov chain model for the SU(2) gauge group, as detailed in our recent study \cite{Boguslavski:2024yto}.
	
	Our results demonstrate that this combined approach stabilizes CL and enables the computation of expectation values for theories where CL previously failed. These findings provide new insights into the conditions under which CL converges correctly and point to pathways for addressing and surpassing its limitations in more complex lattice field theories.
	
	\section{Complexification methods}
	
	In this section, we briefly summarize the two approaches to the numerical sign problem that are applied in this work: complex Langevin and Lefschetz thimbles. The numerical sign problem emerges in the numerical computation of expectation values
	\begin{align}
		\label{eq:expect_values}
		\langle \mathcal{O} \rangle = \frac{1}{Z} \int dx \, \rho(x) \mathcal{O}(x),
	\end{align}
	for observables $\mathcal{O}$, whenever the weight function $\rho(x)=\exp[-S(x)]$ is complex or not positive definite as a consequence of complex actions $S$. These properties prohibit the direct application of standard methods, such as Monte Carlo integration, as the weight function cannot be interpreted as a probability density. 
	
	\subsection{The complex Langevin (CL) method}
	
	The complex Langevin method can be considered an analytical continuation of real Langevin, which, in the context of Euclidean simulations, is sometimes known as stochastic quantization. Instead of directly sampling from the path integral weight $\rho$, which is not applicable when it is complex, we sample from a stochastic process that is described by the complex Langevin equation
	\begin{align}
		\partial_\theta z(\theta) = K(z(\theta)) + \eta(\theta), \quad K(z) = -\partial_z S(z).
	\end{align}
	The first term on the right-hand side is the drift term and is given by the complex derivative of the analytically continued action to the complex plane. The noise term, defining the second term, is the stochastic component of this equation and is characterized by
	\begin{align}
		\langle \eta(\theta) \rangle = 0, \quad
		\langle \eta(\theta) \eta(\theta') \rangle = 2 \delta(\theta - \theta').
	\end{align}
	The process is described by the real probability function $P(z;\theta)$. If the density of the drift magnitude
	\begin{align} \label{eq:pu}
		p(u;\theta) = \int dx dy \, \delta(u - u(x+iy))  P(x+iy;\theta), \quad u(z) = \vert K(z)\vert
	\end{align}
	decays at least exponentially, CL converges correctly. This statement is a criterion of correctness for CL and was formulated by Nagata et al.~in \cite{Nagata:2016vkn}. Another formulation for the criterion is stated by Scherzer et al.~in \cite{Scherzer:2018hid} and requires the absence of boundary terms.
	
	In this context, correct convergence for CL means that the stationary probability density $P(z) = P(z;\theta\to\infty)$ characterizes the complex weight function $\rho$ of the considered system, i.e., if
	\begin{align}
		\int dx dy\, P(x+iy) \mathcal{O}(x+iy) = \frac{1}{Z}\int dx\, \rho(x) \mathcal{O}(x) = \langle \mathcal{O} \rangle_\rho
	\end{align}
	for a sufficient class of holomorphic functions $\mathcal{O}$. In other words, sampling from the complex Langevin process leads to correct expectation values for the considered system.  
	
	\subsection{The Lefschetz thimble (LT) method}
	
	For the LT approach, as for the CL method, we analytically continue the theory to the complex plane to formulate the expectation values as complex line integrals
	\begin{align}
		\langle \mathcal{O} \rangle = \frac{1}{Z} \int_{\mathcal{C}=D} dz \, \rho(z) \mathcal{O}(z).
	\end{align}
	Here the generally complex domain of integration $\mathcal{C}$ coincides with the real domain $D$, i.e., $\mathcal{C}=D\subseteq\mathbb{R}\subseteq\mathbb{C}$ defines a curve in the complex plane. To circumvent the sign problem, this method relies on Cauchy's theorem, which allows us to continuously deform the integration contour while retaining the same integral value. In particular, we define (Lefschetz) thimbles $\mathcal{J}_\sigma$ and anti-thimbles $\mathcal{K}_\sigma$ as solutions of the (anti-)holomorphic flow equations in flow time $t_f$,
	\begin{align}
		\partial_{t_f} z(t_f) = \pm \overline{\partial_z S(z(t_f))}, \quad z(\infty)=z_\sigma.
	\end{align}
	The curves represent the steepest ascent/descent paths for the continued action and asymptotically start in the stationary/critical points $z_\sigma$ indexed by $\sigma$ with $\partial_z S(z_\sigma)=0$. 
	
	Crucially, along these paths, the imaginary part of the action stays constant. Hence, we can rewrite the expectation value as sums of integrals over thimbles that are weighed by a different phase factor
	\begin{align} \label{eq:thimble_integral}
		\langle \mathcal{O} \rangle = 
		\frac{ \sum_\sigma n_\sigma \exp\left[-i S_I(z_\sigma)\right] \int_{\mathcal{J}_\sigma} dz \, \mathcal{O}(z) \exp\left[-S_R(z)\right] }
		{\sum_\sigma n_\sigma \exp\left[-i S_I(z_\sigma)\right] \int_{\mathcal{J}_\sigma} dz \, \exp\left[-S_R(z)\right]},
	\end{align} 
	where $n_{\sigma}$ counts intersections of $\mathcal{K}_\sigma$ with the original domain of integration $D$. Thimbles and their associated stationary points are classified as \emph{relevant} if the intersection number is non-zero, $n_\sigma\neq0$. We emphasize that the determination of intersection numbers and critical points is non-trivial in high-dimensional field theories. In principle, integrating real weight functions circumvents the sign problem. However, a global sign problem arises if many critical points contribute with different phase factors. Additionally, numerical integration over thimbles requires a generally complex Jacobian, reintroducing a softer sign problem than the original.
	
	\subsection{Connecting LT and CL}
	
	In this work, we do not use the LT method to obtain numerical results. Instead, we employ it as a diagnostic tool to identify cases of incorrect convergence in the CL method, thereby reinforcing the connection between both approaches.
	
	When a given theory exhibits exactly one relevant thimble $\mathcal{J}$ we find
	\begin{align}
		\langle \mathcal{O} \rangle = \frac{\int_{\mathcal{J}} dz \, \mathcal{O}(z) \exp\left[-S_R(z)\right]} {\int_{\mathcal{J}} dz \, \exp\left[-S_R(z)\right]},
	\end{align}
	for Eq.~\eqref{eq:thimble_integral}, eliminating the need to compute intersection numbers and phase factors. If multiple thimbles arise due to inherent structures in the complex plane, such as periodicity or point symmetry, we treat them collectively as a single thimble as we can restrict the integration domain accordingly.
	
	At the same time, stationary points play a critical role in the CL process, as they correspond to the points where the CL drift vanishes. Consequently, CL can be regarded as an importance-sampling method \emph{around} the thimbles.
	
	Our findings indicate that CL converges correctly when the system under consideration has exactly one relevant, compact thimble, in agreement with the conjecture proposed in \cite{Salcedo:2016kyy}. Although a formal proof is lacking, we test this hypothesis through numerical simulations on various toy models. These models are manually regularized to allow precise control over the thimble structure.
	
	\subsection{Weight regularizations}
	
	To test the connection between LT and CL in settings where multiple thimbles are relevant to the path integral, we introduce an artificial additive regularization term $R$ for the weight function
	\begin{align}
		\tilde \rho(z) = \rho(z) + R(z;r),
	\end{align}
	where $r\in\mathbb{C}$ controls how dominant the regularization becomes for the considered integrals. To differentiate with respect to which weight function $w$ a certain expectation value is meant, we denote the weight function as a subscript $\langle \cdot \rangle_w$.
	
	We design regularizations such that the resulting effective action,
	\begin{align}
		\tilde S(z) = - \ln \left[ \rho(z) + R(z;r)\right] = S(z) - \ln \left[ 1 + R(z;r) e^{S(z)}\right],
	\end{align}
	exhibit only one relevant compact thimble for which we observe CL to yield correct results. In sections \ref{sec:cosine} and \ref{sec:plm}, we will elaborate on the appropriate construction of regularization terms for different models.
	
	In general, the introduction of $R$ changes the expectation values, and hence, we need to correct for the resulting bias to obtain results for the system we are interested in. An expectation value of the original theory can be expressed in terms of the expectation values with respect to the regularized weight function and the regularization term 
	\begin{align}
		\langle \mathcal{O} \rangle_{\rho} =  \langle \mathcal{O} \rangle_{\tilde \rho} + \mathrm{Bias}_\mathcal{O}[R]  = \langle \mathcal{O} \rangle_{\tilde \rho} + (\langle \mathcal{O} \rangle_{\tilde \rho} - \langle \mathcal{O} \rangle_{R}) \, Q,
		\label{eq:mod_ev_bias}
	\end{align}
	where we introduced the ratio of the partition functions $Q = Z_{R}/Z_\rho$. Note that $r$ and $R$ are chosen such that $\langle \mathcal{O} \rangle_{\tilde \rho}$ and $\langle \mathcal{O} \rangle_{R}$ can be evaluated using complex Langevin. Therefore, the crucial part is the evaluation of $Q$, as the partition functions are not directly accessible. To determine $Q$, we utilize that it is independent of observables. For an observable $\mathcal{O}^*$ for which $\langle \mathcal{O} \rangle_{\rho}=0$, Eq.~\eqref{eq:mod_ev_bias} yields
	\begin{align} \label{eq:ratioQ}
		Q = \left.\langle \mathcal{O}^* \rangle_{\tilde \rho} \right/ \left(\langle \mathcal{O}^* \rangle_{R} - \langle \mathcal{O}^* \rangle_{\tilde \rho}\right).
	\end{align}
	Using the Dyson-Schwinger equation for the unregularized system, we can always construct such observables. We showcase the numerical robustness of this approach for the bias correction in \cite{Boguslavski:2024yto}.
	
	\section{Complex cosine model: well-understood failure of CL} \label{sec:cosine}
	
	We first study the weight regularization approach at the complex cosine model. It is defined by the weight function
	\begin{align}
		\rho(x) = \exp\left[-i \beta \cos(x)\right], \quad x\in[-\pi,\pi].
	\end{align}
	The complex coupling $\beta\in\mathbb{C}$ leads to the sign problem when we consider the numerical evaluation of expectation values in Eq.~\eqref{eq:expect_values}. 
	
	The complex cosine model is intriguing as complex Langevin is known to fail while the expectation values and the stationary probability density $P(x+iy;\theta\to \infty)=(4\pi\cosh^2(y))^{-1}$ of the Langevin process are analytically attainable \cite{Salcedo:2016kyy}. Moreover, CL is known to yield wrong results for this model and it therefore serves as a perfect testing ground for the weight regularization technique.
	
	\begin{figure}[tb]
		\floatbox[{\capbeside\thisfloatsetup{capbesideposition={right,center},capbesidewidth=0.35\textwidth}}]{figure}[\FBwidth]
		{\caption{Density $p(u; \theta \to \infty)$ of the drift magnitude $u$ [Eq.~\eqref{eq:pu}] for the original (blue) and regularized (red) cosine models at $\beta = 0.5$, $r = 0.5$. The original model exhibits power-law decay (green dashed line), violating the convergence criterion, while the regularized model shows exponential decay, ensuring correct CL method convergence.
			}\label{fig:pu_cosine}}
		{\includegraphics[width=0.5\textwidth]{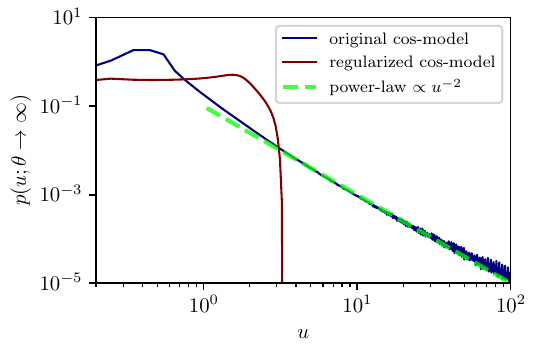}}
	\end{figure}
	
	To address incorrect convergence in this model, we introduce the regularization term
	\begin{align}
		R(z) = r (z^2 - \pi^2) - e^{i \beta}, \label{eq:reg_cosine}
	\end{align}
	adding $-e^{i \beta}$ to impose $\tilde\rho(\pm\pi)=0$ at the integration boundaries. For $\vert r \vert \to \infty$, the regularization pulls the CL process towards the real line, forcing the relevant thimble to connect to these zeros, resulting in a single compact thimble in $z \in [0,\pi]$. Due to point symmetry, the same structure occurs for $z \in [-\pi,0]$. While the regularization breaks periodicity, this is restored by artificially continuing $R(x+iy)=R(x+2\pi+iy)$. The resulting non-holomorphic points at $x=\pm\pi$ do not affect the CL algorithm, as detailed in \cite{Boguslavski:2024yto}.
	
	In Fig.\ref{fig:pu_cosine}, we explicitly examine the criterion of correctness by analyzing the decay of the drift magnitude, as described in Eq.~\eqref{eq:pu}, for the late-time solution of the complex Langevin equation.%
	\footnote{To verify the criterion of correctness independently, we numerically solved the Fokker-Planck equation, which governs the evolution of the probability distribution for the complex Langevin process. The density of the drift magnitude is computed based on this solution.}
	The blue curve demonstrates that, although the probability density of the original model at the coupling $\beta = 0.5$ decays exponentially in the imaginary direction, the drift magnitude exhibits only a power-law decay (blue line). This violates the criterion of correctness. In contrast, for the regularized model at the same coupling, with a regularization force of $r = 0.5$, we observe that $p(u; \theta \to \infty)$ decays exponentially (red line), satisfying the correctness criterion.
	
\begin{figure}[tb]
		\centering
		\includegraphics{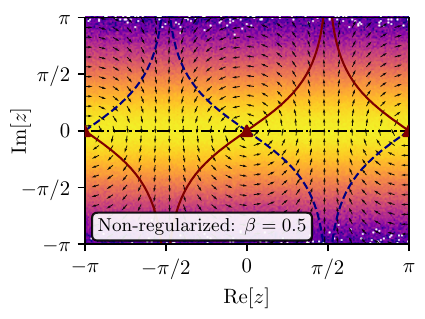}
		\hfill
		\includegraphics{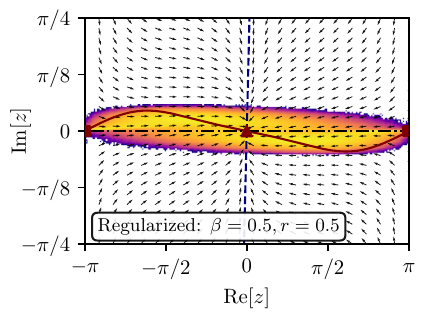}
		\caption{
			Complex Langevin histograms for complex cosine model at the coupling $\beta=0.5$ (\emph{left}) and including the regularization term with $r=0.5$ (\emph{right}). The red-solid and blue-dashed lines represent thimbles and anti-thimbles of critical points (triangles) and connect to singular points of the action (squares). The arrows show the normalized CL drift term for each model. For the non-regularized model, the histogram exhibits a broad structure that is independent of the real part where multiple thimbles are relevant and extend asymptotically to complex infinity. The regularization term in the right panel leads to a sharply defined histogram that indicates a better-conditioned convergence of the CL process.} \label{fig:cosine_histograms}
	\end{figure}

	The impact of the regularization term is evident in the normalized histograms of the complex Langevin process shown in Fig.~\ref{fig:cosine_histograms}. The left panel shows the wide histogram for the non-regularized cosine model at $\beta = 0.5$, reflecting poor convergence. Red solid lines indicate relevant thimbles linked to action stationary points (red triangles), while dashed blue lines show anti-thimbles. With regularization ($r = 0.5$), the right panel shows a sharply localized histogram. A single compact thimble contributes, spanning between the action's singular points (red squares), greatly improving convergence. Details, including bias correction checks, are in \cite{Boguslavski:2024yto}.
	
	\section{Regularization for the SU(2) Polyakov chain model}  \label{sec:plm}
	The second model we discuss in this work is the one-dimensional SU(2) Polyakov chain model, which is described by the action $S[U] = - \beta \,\mathrm{Tr} \left[ P \right]$, with complex coupling $\beta \in \mathbb C$ and where the Polyakov chain $P=U_1 U_2 \cdots U_{N_\mathrm{chain}}$ is given by the product of ${N_\mathrm{chain}}$ links $U_i \in \mathrm{SU(2)}$, $i \in \{1,2,\dots,{N_\mathrm{chain}}\}$.
	
	This model is particularly interesting for our study of weight regularization as the correct and wrong convergence of CL depends on the magnitude of the coupling. Furthermore, we can explicitly study the thimble structure for the model as it allows us to reduce the Haar measure, resulting in a one-parameter model. The reduction leads to the mapping 
	\begin{align}
		\mathrm{Tr}[P] &\to 2 \cos(\phi) \\
		DU \,\exp \left\{ -S[U] \right\} &\to d\phi \, \sin^2(\phi) \exp \left\{ 2 \beta \cos(\phi) \right\} =: d\phi\, J(\phi)\, \rho(\phi),
		\label{eq:Haar_measure_SU2}
	\end{align} 
	where $DU$ denotes the SU(2) Haar measure, $J(\phi)=\sin^2(\phi)$ denotes the Jacobian term for the chosen coordinates and $\phi$ is integrated over $[-\pi, \pi]$. This enables a direct study of the criterion of correctness and its connection with Lefschetz thimbles for an SU(2) theory.    
	
	To achieve correct convergence of CL for this model, we introduce the regularization term $R$,
	\begin{align} \label{eq:su2_red_mod_cos}
		\tilde \rho(\phi) = \rho(\phi) + R(\phi;r) = e^{2 \beta \cos(\phi)} + r ( \cos(\phi) + 1 ).
	\end{align}
	It acts in a similar manner as the regularization term for the complex cosine model. It pulls towards the real line and forces the Lefschetz thimbles to connect to the singularities at the boundary leading to a single relevant thimble structure up to symmetries. Details on the design are discussed in \cite{Boguslavski:2024yto}.
	
	\begin{figure}[tb]
		\floatbox[{\capbeside\thisfloatsetup{capbesideposition={right,center},capbesidewidth=0.35\textwidth}}]{figure}[\FBwidth]
		{\caption{
				Density $p(u; \theta \to \infty)$ of the drift magnitude $u$ for the reduced SU(2) model with $\beta_1 = (1 + \sqrt{3}i)/4$ (teal), $\beta_2 = (1 + \sqrt{3}i)/2$ (blue), and the regularized model with $\beta_2$ and $r = -5$. Exponential decay is observed for $\beta_1$ and regularized $\beta_2$, while $\beta_2$ without regularization exhibits power-law decay (green dashed), satisfying the correctness criterion.}\label{fig:pu_plm}}
		{\includegraphics[width=0.5\textwidth]{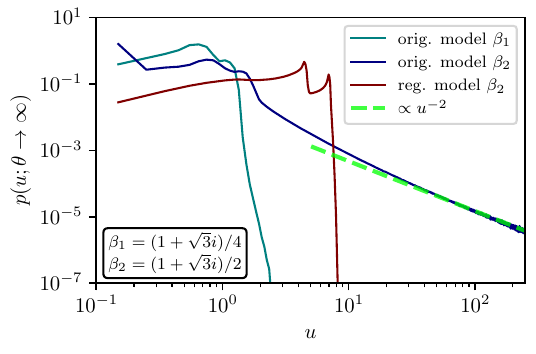}}
	\end{figure}
	
	In Fig.~\ref{fig:pu_plm} we show the drift magnitude at late Langevin times $p(u;\theta\to\infty)$ for the reduced SU(2) Polyakov loop model for two different couplings $\beta_1=(1+\sqrt{3}i)/4$ and $\beta_2=(1+\sqrt{3}i)/2$. For $\beta_1$, the CL process converges correctly as the density of the drift magnitude decays exponentially (teal line). In contrast, the criterion of correctness is not satisfied for $\beta_2$, as indicated by the power-law behavior (blue line). When the regularization term is switched on using $r=-5$, we find that $p(u;\theta\to\infty)$ decays exponentially, and hence, CL yields correct results. The numerical evaluation of expectation values and CL histograms are presented in \cite{Boguslavski:2024yto}. 
	
\begin{figure}[tb]
		\centering
		\includegraphics{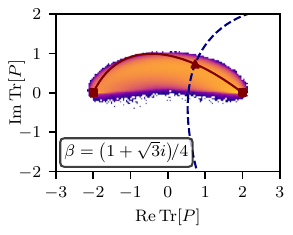}
		\includegraphics{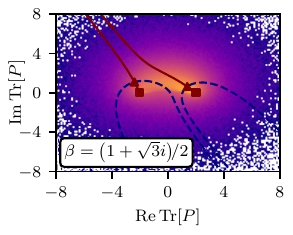}
		\includegraphics{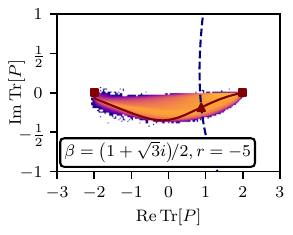}
		\caption{
			Complex Langevin histograms for the SU(2) Polyakov chain model at the coupling $\beta = (1 + \sqrt{3}i)/4$ (\emph{left}), $\beta = (1 + \sqrt{3}i)/2$ (\emph{center}) and for the latter coupling but including the regularization term with $r=-5$ (\emph{right}). The same conventions as in Fig.~\ref{fig:cosine_histograms} are used.} \label{fig:plm_histograms}
	\end{figure}

	We directly run CL simulation for the SU(2) link model and obtain CL histograms for both coupling $\beta_{1,2}$ and the regularized case $\beta_2, r=-5$. Besides imposing the regularization term, we utilize the gauge cooling technique \cite{Seiler:2012wz} to additionally stabilize all our simulations. In Fig.~\ref{fig:plm_histograms}, we show the CL histograms of the trace of the Polyakov loop. For $\beta_1$ (left), the histogram decays rapidly, while for $\beta_2$ (center), it decays slowly, indicating that the CL process undergoes deep excursions into the complex manifold. When the regularization term is employed with $r=-5$, we find a sharp histogram (right). This confirms the notion that the regularization improves the convergence of CL. The numerical evaluation of expectation values and their successful bias correction is discussed in detail in \cite{Boguslavski:2024yto}, where we also extend the study further to SU(3).
	
	\section{Conclusion}
	
	We investigated the correct convergence of complex Langevin (CL) in relation to Lefschetz thimble structures for the complex cosine model and Polyakov chain models with complex couplings for SU(2). Using weight regularization, we showed that CL converges correctly when a single compact thimble contributes to the path integral up to model symmetries. This approach restores convergence and yields accurate expectation values after a bias correction.
	
	The regularization terms were designed to enforce a compact single-thimble structure, and we introduced a method to systematically correct the bias from such regularizations. While additive regularization is effective in lower dimensions, its extension to higher-dimensional theories faces challenges, such as compatibility with lattice models and bias correction tractability. Building on these insights, future work aims to develop kernel transformations that avoid bias correction, as demonstrated in recent studies \cite{Alvestad:2023jgl, Boguslavski:2022dee, Boguslavski:2023unu}.
	
	Our findings emphasize the importance of the thimble structure for CL’s success and motivate further development of stabilization techniques and kernel design, especially for SU(N) field theories, real-time Yang-Mills theory, and finite-density QCD. This work highlights the potential of thimble-guided approaches to overcome longstanding CL failures.

	\begin{acknowledgments}
		This research was funded by the Austrian Science Fund (FWF) under the projects P~34455, P~34764 and \mbox{W~1252}. The computational results presented have been achieved using the Vienna Scientific Cluster (VSC).
	\end{acknowledgments}
	
	\bibliographystyle{JHEP}
	\bibliography{ref}
	
\end{document}